# Large-Scale Optical Reservoir Computing for Spatiotemporal Chaotic Systems Prediction

Mushegh Rafayelyan[,1,*] Jonathan Dong[,1,2] Yongqi Tan[,1] Florent Krzakala[,2] and Sylvain Gigan[1]

[1]*Laboratoire Kastler Brossel, Sorbonne Université, École Normale Supérieure-Paris Sciences et Lettres*
*(PSL) Research University, Centre National de la Recherche Scientifique (CNRS) UMR 8552,*
*Collège de France, 24 rue Lhomond, 75005 Paris, France*

[2]*Laboratoire de Physique de l'École Normale Supérieure, Université Paris Sciences et Lettres (PSL),*
*Centre National de la Recherche Scientifique (CNRS), Sorbonne Université, Université Paris-Diderot,*
*Sorbonne Paris Cité, 24 rue Lhomond, 75005 Paris, France*

Reservoir computing is a relatively recent computational paradigm that originates from a recurrent neural network and is known for its wide range of implementations using different physical technologies. Large reservoirs are very hard to obtain in conventional computers, as both the computation complexity and memory usage grow quadratically. We propose an optical scheme performing reservoir computing over very large networks potentially being able to host several millions of fully connected photonic nodes thanks to its intrinsic properties of parallelism and scalability. Our experimental studies confirm that, in contrast to conventional computers, the computation time of our optical scheme is only linearly dependent on the number of photonic nodes of the network, which is due to electronic overheads, while the optical part of computation remains fully parallel and independent of the reservoir size. To demonstrate the scalability of our optical scheme, we perform for the first time predictions on large spatiotemporal chaotic datasets obtained from the Kuramoto-Sivashinsky equation using optical reservoirs with up to 50 000 optical nodes. Our results are extremely challenging for conventional von Neumann machines, and they significantly advance the state of the art of unconventional reservoir computing approaches, in general.

## I. INTRODUCTION

Recent studies in machine learning show that large neural networks can dramatically improve the network performance; however, their realization with conventional computing technologies is to date a significant challenge. Toward this end, a number of alternative computing approaches have emerged recently. Among them, one of the most studied approaches is reservoir computing (RC). RC is a relatively recent computational framework [1,2] derived from independently proposed recurrent neural network (RNN) models, such as echo state networks (ESNs) [3] and liquid state machines (LSMs) [4]. The main objective of ESNs and LSMs is the significant simplification of the RNN training algorithm by using fixed random injection and fixed internal connectivity matrices. However, it was rapidly understood that the temporally fixed connections allow for the straightforward implementation of RC in optics, electronics, spintronics, mechanics, biology, and other fields [5–12]. Optics is one of the most promising fields to realize large and efficient neural networks due to its intrinsic properties of parallelism, its ability to process the data at the speed of light, and low energy consumption.

There are many interesting approaches to realize photonic reservoir networks based on both time and spatial multiplexing of photonic nodes. The first approach is based on a single nonlinear node with a time-delayed optoelectronic or all-optical feedback in order to get time-multiplexed virtual nodes in the temporal domain [12–24]. Such architectures can reach supercomputer performances, e.g., gigabyte per second data rates for chaotic time-series prediction tasks [25] or million words per second classification for speech recognition tasks [26]. However, their information processing rate is inherently limited, as it is inversely proportional to the number of virtual nodes of the reservoir. Furthermore, a preprocessing of the input information is required, according to the initially defined virtual nodes, which can bring additional complexity to the problem, especially for the large multidimensional inputs. To this end, multichannel delay-based RC architectures consisting of several nonlinear nodes are of special interest [27–31].

---

Another popular approach of photonic RC is based on spatially distributed nonlinear nodes. The latter is endowed by its intrinsic property to process large-scale input information without sacrificing the computation speed. Several theoretical and experimental studies have been performed using on-chip silicon photonics reservoirs consisting of optical waveguides, optical splitters, and optical combiners [32–35]. As reported in Ref. [35], a 16-node reservoir network of modest sizes can reach high information processing bitrates, potentially being able to surpass 100 Gbit s$^{-1}$ in the future. An additional approach toward the spatially extended photonics reservoir is based on a network of vertical-cavity surface-emitting lasers and a standard diffractive optical element (DOE) providing the complex interconnections between the reservoir nodes [36].

Recently, an alternative strategy to spatially scalable photonics reservoir has been introduced based on both liquid crystal spatial light modulators (SLMs) and a digital micromirror device (DMD) [37–41]. In particular, Bueno *et al.* in Ref. [37] demonstrate a reservoir network of up to 2500 diffractively coupled photonic nodes using a liquid crystal SLM coupled with a DOE and a camera. The input and output information in their network is provided via single nodes. This last limitation is waived by Dong *et al.* in Ref. [38] using a DMD to encode both the reservoir and the input information through the binary intensity modulation of the light. Later, Dong *et al.* in Ref. [39] implemented the same approach to get large-scale optical reservoir networks using a phase-only SLM that could provide an 8-bit encoding of the reservoir and the input information through the spatial phase profile of the light instead of the former binary encoding option. We stress that the key element in both aforementioned optical networks is the strongly scattering medium that guarantees random coupling weights of a very large number of photonic nodes and their parallel processing. Such networks practically can host as many nodes as the number of pixels provided by the DMD and the camera [42,43].

In this work, we exploit the potential of the platform provided by Refs. [38,39] to extend our recent achievements toward multidimensional large chaotic systems predictions. Accordingly, we report on the first experimental realization of the recently introduced state-of-the-art benchmark test [44], performing recursive predictions on the Kuramoto-Sivashinsky (KS) chaotic systems. To highlight the scalability of our approach, we measure the computation time of similar reservoir networks provided either by a high-end conventional computer or by our optical scheme. In contrast to conventional computers, where the time of the computation scales quadratically with the size of the network, the computation time of our optical scheme is almost independent of the number of photonic nodes. More precisely, we observe a relatively mild linear dependence due to electronic overheads, while the optical computation remains fully parallel and independent of the reservoir size. Our results are hardly reachable by the conventional von Neumann machines, and they significantly advance the state of the art of the unconventional reservoir computing approaches, in general.

## II. CONVENTIONAL RESERVOIR COMPUTING

We now briefly introduce the concept of conventional RC. An input vector $\mathbf{i}(t)$ of dimension $D_{in}$ is injected to a high-dimensional dynamical system called the "reservoir" [see Fig. 1(a)]. The reservoir is described by a vector $\mathbf{r}(t)$ of dimension $D_{res}$ that is the number of reservoir nodes. The initial state of the reservoir is defined randomly. Let $\mathbf{W}_{res}$ matrix define the internal connections of the reservoir nodes and $\mathbf{W}_{in}$ matrix define the connections between the input and the reservoir nodes. Both matrices are initialized randomly and fixed during the whole RC process. The state of each reservoir node is a scalar $r_j(t)$, which evolves according to the following recursive relation:

$$\mathbf{r}(t + \Delta t) = f[\mathbf{W}_{in}\mathbf{i}(t) + \mathbf{W}_{res}\mathbf{r}(t)], \qquad (1)$$

where $\Delta t$ is the discrete time step of the input and $f$ is an elementwise nonlinear function. According to Eq. (1), the reservoir is defined as a high-dimensional dynamical system endowed with a unique memory property; namely, each consequent state of the reservoir contains some exponentially decaying information about its previous states and about the inputs injected until that moment. Interestingly, the memory capacity of the reservoir is mainly defined by the number of reservoir nodes and the nonlinear activation function $f$.

During the training phase, the input $\mathbf{i}(t)$, defined in the time interval $-T \leq t \leq 0$, is fed to the reservoir, and the corresponding reservoir states are recursively calculated. The final step of the information processing is to perform a simple linear regression that adjusts the $\mathbf{W}_{out}$ weights so that their linear combination with the calculated reservoir states makes the actual output $\mathbf{o}(t)$ to be as close as possible to the desired output $\tilde{\mathbf{o}}(t)$:

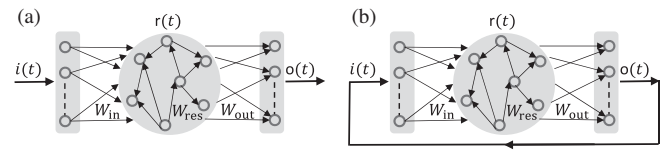

FIG. 1. The sketch of the conventional reservoir computing paradigm in (a) training and (b) predicting phases. The vectors $\mathbf{i}(t)$, $\mathbf{r}(t)$, and $\mathbf{o}(t)$ describe the injected input, the corresponding reservoir states, and the trained output, respectively. All three layers of the network are described by $\mathbf{W}_{in}$, $\mathbf{W}_{res}$, and $\mathbf{W}_{out}$ interconnection matrices. The first two are initialized randomly and are held fixed throughout the whole computation process, while the last one is trained by linear regression. In the prediction phase, the feedback loop from the predicted output defines the next injected input.

$$\text{RMSE} = \sqrt{\frac{1}{D_{\text{out}}T}\sum_{t=-T}^{0}\|\tilde{\mathbf{o}}(t)-\mathbf{o}(t)\|^2}, \qquad (2)$$

where

$$\mathbf{o}(t) = \mathbf{W}_{\text{out}} \cdot \mathbf{r}(t), \qquad (3)$$

$$\mathbf{W}_{\text{out}} = \text{argmin}(\text{RMSE}). \qquad (4)$$

RMSE is the root mean square error, and $D_{\text{out}}$ is the number of the output nodes, i.e., the dimension of the vector $\mathbf{o}(t)$. An additional regularization term $\lambda\|\mathbf{W}_{\text{out}}\|^2$ ($\lambda$ is a scalar) can be used to find the solution of Eq. (4) to avoid overfitting, especially when the number of reservoir nodes is larger than the number of training examples. Note that the output weights are the only parameters that are modified during the training. The random input and reservoir weights are fixed throughout the whole computational process, and they are used to randomly project the input into a high-dimensional space, which increases the linear separability of inputs.

In order to perform predictions about the future evolution of $\mathbf{i}(t)$ ($t > 0$) using the calculated reservoir states $\mathbf{r}(t)$ in $-T \le t \le 0$, one needs to train the output weights $\mathbf{W}_{\text{out}}$ to predict the next time step of the input, namely, $\tilde{\mathbf{o}}(t) = \mathbf{i}(t + \Delta t)$. Afterward, the future evolution of $\mathbf{i}(t)$ for $t > 0$ can be predicted by replacing the input by the subsequent prediction $\mathbf{o}(t)$, as shown in Fig. 1(b). Consequently, during the prediction, the reservoir evolves step by step, by replacing the subsequent input with the last prediction every time.

## III. OPTICAL RESERVOIR COMPUTING

The experimental setup to perform the optical RC is shown in Fig. 2 and detailed in the Appendix A. The key optical components in the setup are the phase-only SLM, the scattering medium, and the camera. The SLM provides both the encoding of the input vector $\mathbf{i}(t)$ of dimension $D_{\text{in}}$ and the encoding of the subsequent reservoir state $\mathbf{r}(t)$ of dimension $D_{\text{res}}$ (total dimension $D_{\text{in}} + D_{\text{res}}$) into the phase spatial profile of the light. The scattering medium ensures their random linear mixing, which is equivalent to their linear multiplications with large dense random matrices consisting of independent and identically distributed (i.i.d.) random complex variables [45,46] (see more details about light scattering in Appendix B). Finally, the camera performs a nonlinear readout of the complex field intensity for the next reservoir state $\mathbf{r}(t + \Delta t)$, that is sent back by the computer to the SLM in order to be displayed with new input, and the process repeats. The upper and the lower insets in Fig. 2 are respective examples of images displayed on the SLM and detected by the camera.

There are a number of tunable parameters regarding the encoding of the input and the reservoir states onto the SLM

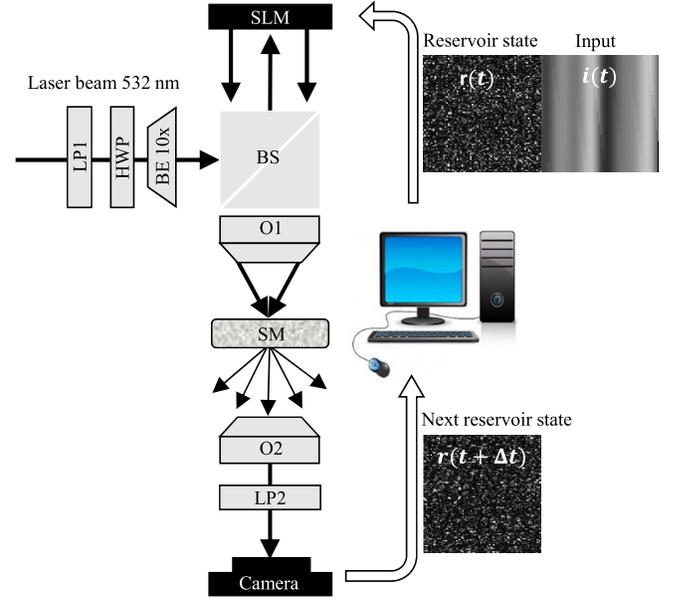

FIG. 2. Experimental setup to perform an optical reservoir computing. The SLM receives from the computer the consequent input $\mathbf{i}(t)$ concatenated with the reservoir state $\mathbf{r}(t)$ and imprints it into the spatial phase profile of the reflected beam (see the upper inset as a typical example). The scattering medium (SM) provides a complex linear mixing of the whole encoded information. Finally, the camera performs a nonlinear readout for the next reservoir state $\mathbf{r}(t + \Delta t)$ (see the lower inset as a typical example), which is sent by the computer back to the SLM to be displayed with new input, and the process repeats. LP1 and LP2, linear polarizers; HWP, half-wave plate; BE, beam expander; BS, beam splitter; O1 and O2, objectives.

that we describe here. Without loss of generality, we assume that the number of gray levels of the camera and the SLM are equal to 256. The SLM is calibrated such that the gray levels from 0 to 255 linearly map to phase delays of 0 to $2\pi$. Furthermore, we assume that the whole input dataset is initially scaled from 0 to 255 and the acquisition time of the camera is initially adjusted to provide unsaturated reservoir states again ranging from 0 to 255. Accordingly, the encoding of the input and the reservoir states onto the SLM can be described by $\mathbf{i}(t) \to s_{\text{in}}\mathbf{i}(t)$ and $\mathbf{r}(t) \to s_{\text{res}}\mathbf{r}(t)$, respectively, with two scaling factors $0 \le s_{\text{in/res}} \le 1$. Additionally, each scalar value from the input and reservoir states can be encoded into multiple number of SLM pixels forming a macropixel. These modifications are performed in the computer every time before sending the input and reservoir states to the SLM, since we obtain their explicit forms for time $t$ only at $t - \Delta t$, except the input states for training which are already known before starting the training and can be modified a priori. We note that the mentioned modifications require only simple multiplication with scalar values and do not impose a large computational overhead. The number of pixels in one macropixel is denoted by $p_{\text{in}}$ for the input encoding and $p_{res}$ for the reservoir states encoding. Accordingly, the reservoir computing in our optical scheme can be described by the following recursive relation:

$$\mathbf{r}(t + \Delta t) = F[s_{res}\mathbf{r}(t) \oplus \mathbf{J}_{p_{res}}, s_{in}\mathbf{i}(t) \oplus \mathbf{J}_{p_{in}}], \quad (5)$$

where the function $F$ stands for the whole optical setup; i.e., it takes the encoded matrices corresponding to the input and the reservoir state as two arguments, sends them to the SLM, and returns the next reservoir state detected by the camera. The symbol $\oplus$ refers to the Kronecker product, and $\mathbf{J}_{p_{in/res}}$ refers to the all-ones matrix with $p_{in/res}$ number of rows and columns in order to ensure the macropixel encoding of the SLM.

In order to get a more detailed description of our optical scheme, we also provide a mathematical relation that models the light propagation and the consequent RC with well-known mathematical functions:

$$\mathbf{r}(t + \Delta t) = f\{\mathbf{W}_{res}g[\mathbf{r}(t)] + \mathbf{W}_{in}g[\mathbf{i}(t)]\}, \quad (6)$$

where $\mathbf{W}_{res}$ and $\mathbf{W}_{in}$ are random dense matrices describing the scattering of the light in the setup. $f$ and $g$ are nonlinear functions associated with the intensity readout by the camera and the phase encoding by the SLM, respectively. Namely, for a vector $\mathbf{q} = [q_1, q_2, ...]^T$, $f(\mathbf{q}) = [|q_1|^2, |q_2|^2, ...]^T$ and $g(\mathbf{q}) = [\exp(i\pi s q_1), \exp(i\pi s q_2), ...]^T$ with $0 \leq s \leq 2$. Note that all above mentioned operations are implicitly included in the function $F$ in Eq. (5).

The mathematical framework describing our optical network is very similar to the conventional RC network provided by Eq. (1). The main difference is that an additional complex exponent function is applied in Eq. (6) to account for the phase encoding of the SLM, which changes the overall nonlinearity in the recursive relation. One can also note that $\mathbf{W}_{res}$ and $\mathbf{W}_{in}$ are complex-valued matrices here in contrast to the conventional RC, where the connection matrices are real valued. Accordingly, Eqs. (5) and (6) together give the whole picture of information processing in our optical scheme.

During the training phase, as soon as the reservoir states for the given time interval $-T \leq t \leq 0$ are optically calculated, a simple linear regression is executed in the conventional computer to adjust the $\mathbf{W}_{out}$ weights such that their linear combination with the calculated reservoir states makes the actual output to be as close as possible to the next time step of the input $\mathbf{i}(t + \Delta t)$ [see Eqs. (2)–(4)]. Finally, to predict the future evolution of $\mathbf{i}(t)$ for $t > 0$, we make a feedback loop from the output to the input by replacing the next input $\mathbf{i}(t + \Delta t)$ on the SLM with the one-step prediction $\mathbf{W}_{out}\mathbf{r}(t)$, as is done in conventional RC in Fig. 1(b).

In general, the RC and its different optical implementations have proven to be very successful for various tasks, such as spoken digits recognition, temporal Exclusive OR task, Santa Fe, Mackey-Glass, or Nonlinear Autoregressive Moving Average time-series prediction [5,9,11,13,17,27,47]. Recently, Pathak *et al.* [44,48] proposed a new state-of-the-art benchmark test performing predictions on KS spatiotemporal chaotic datasets with the conventional RC (see more details about the KS equation in Appendix C). In the next section, we use the optical RC setup in Fig. 2 to predict the dynamical evolution of KS spatiotemporal chaotic systems.

## IV. EXPERIMENTAL RESULTS

Initially, we apply the optical RC scheme on the spatiotemporal KS datasets with a similar set of parameters as reported in Ref. [44]. Namely, the spatial domain size $L$ of the scalar field $u(x, t)$ is $L = 22$ in the KS equation [see Eq. (C1) in Appendix C], which is integrated on the grid of $N_x = 64$ equally spaced spatial points and $N_t = 90500$ equally spaced time steps with $\Delta t = 0.25$ (see more details in Appendix C). The first $9 \times 10^5$ time steps of the dataset are used to train the optical reservoir, while the remaining 500 time steps are kept in order to be compared with predicted data. The input and reservoir sizes are $D_{in} = 64$ and $D_{res} = 10^4$, respectively.

In general, it is believed that the optimum prediction performance of RC schemes is reached when the reservoir computer parameters are tuned to the edge of chaos [49]. Accordingly, before starting the actual experiment, we perform a grid search to optimize a set of tunable parameters in our optical scheme. It turns out that the optimal prediction performance is observed when $s_{res} = s_{in} = 0.5$; i.e., the input and reservoir states are encoded between 0 and 128, thus providing a phase modulation of the light from 0 to $\pi$. Furthermore, the macropixel sizes are taken $p_{res} = 64$ and $p_{in} = 10\,000$ to ensure equal importance ratios between the input and reservoir states encoded on the SLM. Consequently, during the RC process, the total number of pixels occupied on the SLM by the input and the reservoir states together is equal to $p_{res}D_{res} + p_{in}D_{in} = 128 \times 10^4$. We also apply a slight regularization with $\lambda = 0.07$ during the linear regression process [see Eqs. (2)–(4)]. Noteworthy, the nonlinear activation function provided by the camera intensity readout may easily be further tuned by both digital and analog ways and may substantially increase the performance of the optical network (see further details in Appendix D). Figure 3 shows an example of the true KS dataset [see Fig. 3(a)], the corresponding prediction [see Fig. 3(b)], and their difference [see Fig. 3(c)]. We observe that the optical reservoir network can predict with excellent accuracy the dynamical change of the KS dataset up to two Lyapunov time. Lyapunov time is a characteristic quantity of dynamical chaotic systems defining the minimum amount of the time for two infinitesimally close states of the system to diverge by a factor of $e$. The latter is defined by the largest Lyapunov exponent $\Lambda_{max}$, and in this particular case $\Lambda_{max} = 0.043$ (see

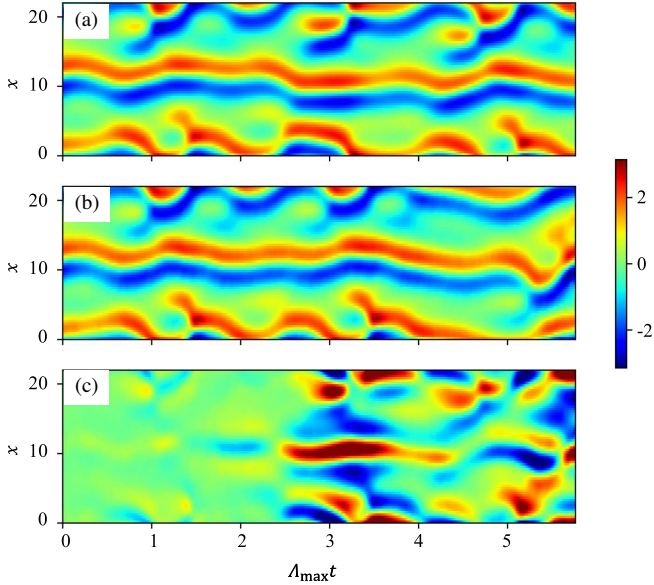

FIG. 3. Experimental Kuramoto-Sivashinsky spatiotemporal chaotic datasets prediction by optical reservoir computing. The spatial domain size of the chaotic system is $L = 22$. The number of the photonic nodes in the reservoir is $D_{res} = 10^4$. (a) Actual data. (b) Reservoir prediction. (c) Error: (a) minus (b). $t = 0$ corresponds to the start of the prediction in the test phase. Each unit on the temporal axis represents the Lyapunov time defined by the largest Lyapunov exponent $\Lambda_{max}$ and detailed in Appendix C.

Appendix C and Table I). Furthermore, for quantitative analyses, we repeat the same experiment of Fig. 3 for 100 different sets of training and testing datasets. The RMSE values for each testing sample are calculated and normalized according to the standard deviation of target output. Figure 4 shows the NRMSE dependencies for each testing sample [see Fig. 4(a)] and the mean NRMSE curve averaged over all 100 samples [see Fig. 4(b)]. We note that the prediction performance varies significantly depending on the test sample, as seen from Fig. 4(a). This effect is related to the RC algorithm, in general, which is addressed in Ref. [50].

Although the prediction results of Figs. 3 and 4 indicate the potential of the optical RC to predict large spatiotemporal chaos, we emphasize that, for larger sizes of the problem, i.e., for larger values of $L$, in order to get qualitatively similar prediction performances, one needs to increase the size of the reservoir. To this end, we perform experiments applying the same reservoir network hosting $D_{res} = 10^4$ photonic nodes on KS datasets with the spatial sizes of $L = 12, 22, 36, 60,$ and $100$. As seen in Fig. 5(a), the prediction performance of optical RC decreases rapidly as the system size $L$ increases. On the other hand, for the given KS dataset of spatial size $L = 60$, Fig. 5(b) shows that the prediction performance of our optical scheme is recovered back by increasing the size of the network. We note that large reservoirs mostly improve the prediction performance in the intermediate temporal regions where the prediction is neither too simple nor too difficult. In both plots, the temporal axis is normalized according to the $\Lambda_{max} = 0.043$ corresponding to $L = 22$; however, we note that the value of the largest Lyapunov exponent is dependent on the spatial domain size $L$ of the system (see Table I in Appendix C). Finally, the different reservoir dimensions in Fig. 5(a) imply different macropixel sizes of encoding in order to maintain the same overall encoding number of pixels on the SLM corresponding to the reservoir states.

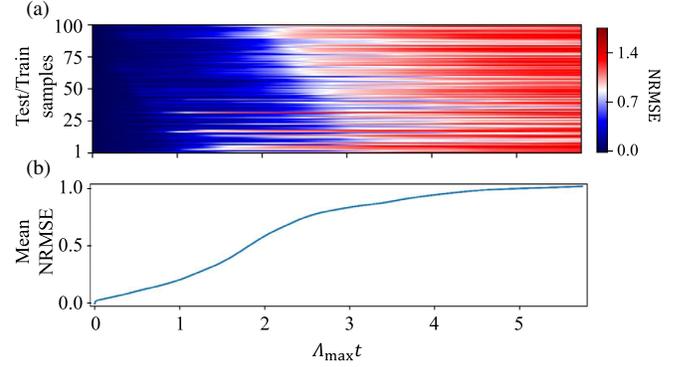

FIG. 4. (a) Normalized root mean square errors (NRMSE) calculated for 100 sets of training and testing KS datasets having the same parameters of the problem as in Fig. 3. (b) The mean NRMSE as a result of averaging (a) along its vertical axis.

Note that the realization of large reservoir networks in conventional computers is not an easy task, since the computation time and the operative memory grow quadratically with respect to the number of network nodes. Therefore, Pathak et al. propose in Ref. [44] a new approach realizing distributed computing on a large set of parallel reservoirs of moderate sizes, each of which hosts 5000 neurons and predicts a local region of the spatiotemporal chaos. Although the parallel reservoir approach works well for predicting KS systems, it is valid if only local spatial nodes in the system get coupled at each time step. In other words, the total effective reservoir—that consists of these parallel reservoirs—provides nonzero connections only close to the diagonal region. The latter can be crucial for more complicated problems where fully connected spatial nodes might be required, such as many-body problems. In this context, our optical scheme allowing fast and large optical reservoirs with nonzero and all-to-all randomly coupled optical nodes may perform better.

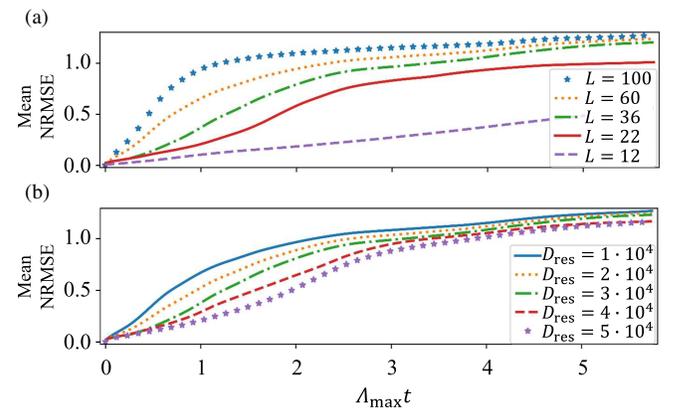

FIG. 5. (a) The mean NRMSE in the predictions of the KS system as a function of time using the same optical network as in Figs. 3 and 4 but for different system sizes $L = 12, 22, 36, 60,$ and $100$. (b) For the case of $L = 60$, we observe improvement of the prediction performance as the number of photonic nodes in reservoir increases from $D_{res} = 10^4$ to $D_{res} = 5 \times 10^4$.

To test the scalability of our approach, we perform a number of experiments on our optical scheme for different reservoir sizes and record the average time of the reservoir updating process. We use the same parameters of the problem as in Fig. 3 but without applying a Kronecker product in Eq. (5), since $p_{res} = p_{in} = 1$; i.e., each pixel of the SLM is one node in the optical network. However, the computation time accounts for all the other conventional computing resources that are needed to our optical setup to calculate the reservoir states, such as the concatenation and rescaling. In order to make a thorough comparison, we perform numerical computations of reservoir states for the same reservoir sizes using both state-of-the-art CPU and GPU technologies (see their characteristics in Refs. [51,52]). Figure 6 shows that the optical RC is slower only for small reservoir sizes. The situation changes rapidly for large network sizes, since the computation time of conventional RC approaches based on CPU and GPU technologies grows quadratically, while optical RC scales only with a mild linear dependence with respect to the number of nodes of the reservoir. Hence, for large reservoir sizes, our optical network is much faster than conventional reservoir computers. Noteworthy, the optical computation in our setup is inherently parallel, and the linear slope is due only to the limited communication bandwidth from the camera to the SLM. As another crucial advantage, large reservoirs require tremendous capacities of operating memory from the conventional computers to store the large random connection matrices $W_{res}$ and $W_{in}$ (see the memory limits region in the inset in Fig. 6), while our optical scheme can leverage a large number of photonic nodes without using large operating memory. However, the limited stability time of our optical scheme does not allow us to use such a large number of photonic nodes in practical problems, e.g., the prediction of the KS systems, since the large number of photonic nodes implies a longer computation time of one reservoir state in our optical scheme. We emphasize that further improvements of the stability time of our optical scheme will allow one to implement reservoirs with larger capacities than 50 000 photonic nodes used in Fig. 5. Furthermore, faster SLMs and cameras are available that can considerably lower the absolute time of computation in our optical scheme while maintaining its linear dependence on the size of the reservoir (see more information about the SLM and camera used in our setup in Appendix A). As both optical and conventional reservoir computing approaches use the same conventional linear regression procedures and only at the end of the training phase, it requires relatively small computation time compared to the reservoir state calculation process. Accordingly, to clearly distinguish the advantage that our optical approach brings into reservoir computation, we do not include the linear regression in Fig. 6 measurements.

Finally, we stress that the advantage of our optical scheme over other optical realizations is not only due to the possibility of using a large number of pixels from the camera and SLM as nodes in the optical network. An important advantage lies in using the complexity of the multiply scattering medium, that ensures all-to-all random mixing of millions of SLM modes with millions of CCD pixels to reach such large network sizes. Although some correlations may be present when measuring all input-output channels, in particular, due to the so-called bimodal distribution of eigenvalues [42], these correlations wash out very quickly when measuring a partial transmission matrix, for instance, with a nonideal numerical aperture, or measuring only one polarization, and one quickly retrieves the results predicted for uncorrelated i.i.d random matrix theory [53,54]. Other indirect experiments probing the fact that strongly scattering media have a large number of uncorrelated modes can be found, for instance, in Ref. [43], where the authors achieve light focusing through the scattering medium with an unprecedented enhancement factor greater than $10^5$, with more than 1 million controlled modes, showing no apparent limitation by correlation of the medium. We also note the work of Keriven, Garreau, and Poli [55], where the authors achieve $10^5$ independent modes using a similar optical architecture as in our case. In all these experiments, the main limitation is the measurement time, and there is no indication that the number of connections would saturate. Consequently, we believe that optical reservoirs using strongly scattering media can be scaled to millions of nodes in the future. Finally, we note that relatively large network sizes are also reachable using periodic diffraction gratings; for instance, the possibility to reach up to 30 000 nodes is claimed in Ref. [56], however, without supporting all-to-all random connectivity between photonic modes.

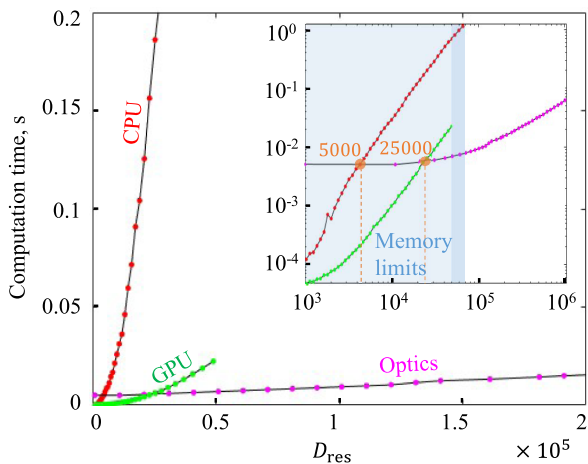

FIG. 6. Computation time of one reservoir state for different reservoir sizes performed on high-end CPU, GPU, and the proposed optical scheme. The inset shows the performance curves in a logarithmic scale extended for larger reservoir sizes. The turning points where the optical scheme starts to perform faster than the CPU and GPU correspond to $D_{res} \approx 5000$ and $D_{res} \approx 25000$, respectively. Note that the largest reservoir size used in this work for the prediction task is 50 000 (see Fig. 5), which is close to the CPU and GPU memory limits and where our implementation is about 4 times faster than the GPU and 100 times faster than the CPU.

## V. DISCUSSION AND CONCLUSION

To measure the computing performances our simple setup can reach, we can estimate the average number of operations per second performed during the process of the RC. As a rough estimate, the optical scheme we propose can host $10^6$ photonic nodes in the network (limited by the pixel numbers on SLM and CCD, respectively). One iteration of the network thus corresponds to about $10^{12}$ trivial mathematical operations in Eq. (1), such as multiplication, sum, etc. Assuming that the SLM and the camera have typical speeds of 100 Hz, our optical setup performs in the order of $10^{14}$ OPS (operations per second). This result is not far from the current state of the art of supercomputers, which ranges from $10^{15}$ to $10^{17}$ OPS. Consequently, without significant energy consumption or a large number of processors, our optical setup can perform a RC close to the performances of the state-of-the-art supercomputer technologies. Note that similar calculations have been performed using the optical processing unit developed by LightOn, with a different modulation scheme (binary amplitude modulation), in Refs. [38,57].

Although light propagation in our optical setup provides fully parallel information processing, Fig. 6 shows that the electronic feedback from the camera to SLM is a bottleneck resulting in a slight linear growth of the overall computation time as the number of nodes increases. One way to overcome this might be the use of the field-programmable gate arrays (FPGAs) instead of the computer to provide the information transfer in much larger bandwidths. Furthermore, FPGAs contain an array of programmable logic blocks that can be configured to apply a given complex operation on the data transferred from the camera to the SLM. Another approach that can impact the overall computation speed is based on nonlinear light-matter interactions, where the naturally generated response from the matter can be used as feedback of the RC network [58–62].

In conclusion, we propose an optical reservoir computing network that can perform, for the first time to our knowledge, predictions on large multidimensional chaotic datasets. We use the Kuramoto-Sivashinsky equation as an example of a spatiotemporal chaotic system. Our predictions on the chaotic systems of large spatial sizes confirm that in order to have comparable prediction performances one has to increase the optical network sizes, too. Finally, we experimentally demonstrate that our optical network can be scaled to a million nodes. Its computation time grows linearly only when the number of nodes increases, due to electronic overheads, while the speed of the optical part (the matrix multiplication) is independent of the reservoir size and does not require any memory storage. Our results, that are very hard to achieve by conventional von Neumann machines, open the prospect to achieve predictions on very large datasets of practical interest, such as turbulence, at high speed, and low energy consumption.

## VI. CODE AND DATA AVAILABILITY

All data and codes can be accessed at Ref. [63].

## ACKNOWLEDGMENTS

We acknowledge funding from the Defense Advanced Research Projects Agency (DARPA) under Agreement No. HR00111890042. S. G. and J. D. also acknowledge partial support from H2020 European Research Council (ERC) (Grant No. 724473).

## APPENDIX A: EXPERIMENTAL SETUP

The laser beam with 532 nm wavelength is expanded using a 10× beam expander (BE). The linear polarizer LP1 and the half-wave plate (HWP) are used to polarize the light parallel to the extraordinary axis of the liquid crystal SLM to ensure a pure phase shaping of the light. The SLM receives from the computer the consequent input information $\mathbf{i}(t)$ concatenated with the reservoir state $\mathbf{r}(t)$ at the given moment and imprints it into the phase spatial profile of the reflected beam. The light propagates further through the first objective O1 with 10× optical magnification and numerical aperture NA = 0.1. Furthermore, the light gets focused on the strongly scattering medium (SM) with 0.5 mm thickness and collected by the second objective O2 with 20× magnification and NA = 0.4. The resulted intensity speckle pattern is detected by the CMOS camera through a crossed linear polarizer LP2 with respect to the initial polarization of the beam in order to enhance the contrast of detected speckle pattern. Finally, the camera sends back to the computer the detected speckle pattern as a new state of the reservoir, that is going to be displayed on the SLM with new input, and the process repeats. We use in our experimental setup a liquid crystal SLM from Meadowlark Optics (Model No. HSP192-532) and CMOS camera from Basler (Model No. acA2040-55um), respectively, having 1920 × 1152 and 2048 × 1536 spatially distributed pixels and, respectively, providing ∼50 and 64 Hz speeds at fully functioning regimes.

## APPENDIX B: LIGHT SCATTERING

When light encounters refractive index inhomogeneities, it gets scattered, modifying its direction of propagation. Light scattering through the thick scattering medium is a complex process accompanied by a large number of scattering events resulting in a speckle pattern at the exit of the scattering medium which is the total interference between all complex scattering paths. Thanks to a large number of scattering events, the speckle image is seemingly random, and its statistical properties are well characterized [64]. It represents a signature of the particular disordered medium and for a given incident field is different from one scattering sample to another.

Light propagation through the scattering medium still is a linear process. Therefore, the output over a set of detectors for the given set of input sources can be described as the product between the incident electric field and the transmission matrix (TM). So, the TM is a characteristic for the particular setup including the input sources, output detectors, and all the optical elements with the scattering medium used inside the setup. As shown in Refs. [42,46], the TM is a dense random matrix when a thick disordered medium is placed between the SLM and the camera, and it can be measured experimentally. Nowadays, SLMs and cameras based on silicon photonics can afford a few millions of pixels; thus, the TM in conventional computers can reach gigantic sizes. We cannot possibly hope to measure such a large matrix, as it requires very large computation time, and it is impossible to store it in the memory of a computer. However, we can leverage the very large dimensionality of TM without measuring it by using well-developed algorithms where the explicit form of the TM is not required [45]. One of those algorithms is RC, which requires large random matrices held fixed throughout the whole computation process.

## APPENDIX C: KURAMOTO-SIVASHINSKY TIME SERIES

The KS equation is a model of nonlinear partial differential equation frequently encountered in the study of nonlinear chaotic systems with intrinsic instabilities, such as the velocity of laminar flame front instabilities or the hydrodynamic turbulence [65]. The one-dimensional KS partial differential equation is

$$u_t = -uu_x - u_{xx} - u_{xxxx}, \tag{C1}$$

where we assume the scalar field $u = u(x, t)$ is periodic with $L$, $u(x + L, t) = u(x, t)$, and, thus, the solution is defined in the interval $[0, L)$. Note that the dimension of the attractor is defined by the value of $L$, and the dependence is linear for large values of $L$. We integrate Eq. (C1) with periodic boundary conditions on a grid of $Q = 64$ equally spaced spatial points with $\Delta t = 0.25$ time step as in

TABLE I. The largest Lyapunov exponent for different spatial domain sizes.

| $L$ | 12 | 22 | 36 | 60 | 100 |
|---|---|---|---|---|---|
| $\Lambda_{max}$ | 0.003 | 0.043 | 0.080 | 0.089 | 0.088 |

Ref. [44] (see Fig. 4.2 from Ref. [66] for the MATLAB code to solve the KS equation). As an initial condition for $u(x, 0)$, we take a random vector with elements varying between 0 and 0.25. The obtained solution contains $Q$ time series, which we denote by the vector $\mathbf{u}(t)$ and use as the reservoir input.

The dynamics of chaotic systems can be described by a quantity called the Lyapunov exponent that measures the exponential divergence of initially close trajectories in the phase space of the system in which all possible states of a system are represented as unique points. As is known, the spatial domain size $L$ of the KS system strongly affects its dynamics, thus changing the corresponding largest Lyapunov exponent. We provide in Table I the $\Lambda_{max}$ values for typical domain sizes as measured in Ref. [67].

## APPENDIX D: NONLINEAR ACTIVATION FUNCTION IN THE OPTICAL RESERVOIR COMPUTING SCHEME

Nonlinear activation function of the optical reservoir can be tuned using both digital and analog approaches. To implement an analog nonlinearity, one can tune the camera gain or exposure time above the saturation level resulting an additional analog inverted ReLU-like nonlinear function applied to the output intensity image. Moreover, it is also possible to modify the look-up table of the camera to engineer any fixed analog nonlinear function applied to the intensity image. Both these methods are done at the hardware level without any computational overhead. The digital approach is easier to implement but requires additional computation power. Accordingly, one needs to apply the nonlinear function numerically on the detected intensity image before sending it back to the SLM. All approaches may improve the performance of such optical RC but require thorough grid searches to find the appropriate nonlinear function and tune its parameters.

To demonstrate the potential of the additional nonlinearity in such optical schemes, we perform simulations of conventional RC [see Eq. (1)] on a similar KS dataset presented in Fig. 3 both for the standard intensity readout, $f(\mathbf{q}) = |\mathbf{q}|^2$, and for an additional nonlinear function applied on it, $f(\mathbf{q}) = 1 - \tanh(2.06|\mathbf{q}|^2)$; see Fig. 7. The hyperbolic tangent nonlinearity with appropriately tuned parameters improves the prediction performance by approximately 50%. The state of the reservoir at $t = 0$ for both activation functions is initialized by a random normal distribution with a mean value of 0.5 and with a standard deviation of 0.2. The interconnection matrices $\mathbf{W}_{in}$ and $\mathbf{W}_{in}$ for both activation functions are initialized with a zero mean value, while their standard deviations for the intensity readout are 0.78 and 0.41, respectively, and for the hyperbolic tangent activation function are 0.9 and 0.54, respectively. All the parameters are monitored throughout the grid search to optimize the prediction performance.

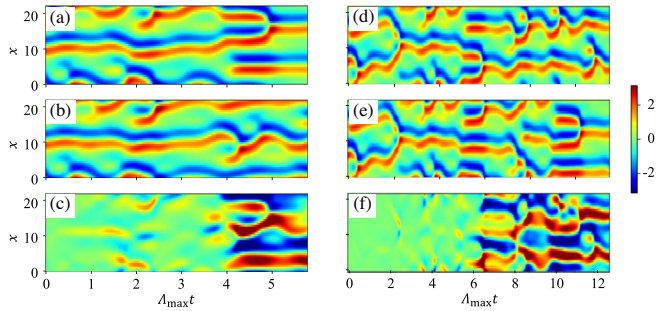

FIG. 7. Kuramoto-Sivashinsky spatiotemporal chaotic datasets prediction by conventional reservoir computing that models (a)–(c) the standard intensity readout activation function and (d)–(f) the appropriately tuned hyperbolic tangent nonlinear function applied on it. (a),(d) Actual datasets. (b),(e) Reservoir predictions. (c),(f) Errors. The spatial domain size of the chaotic system and the number of the photonic nodes in the reservoir are the same as in Fig. 3.

[1] D. Verstraeten, B. Schrauwen, M. dHaene, and D. Stroobandt, *An Experimental Unification of Reservoir Computing Methods*, Neural Netw. **20**, 391 (2007).

[2] M. Lukoševičius and H. Jaeger, *Reservoir Computing Approaches to Recurrent Neural Network Training*, Comput. Sci. Rev. **3**, 127 (2009).

[3] H. Jaeger, *The Echo State Approach to Analysing and Training Recurrent Neural Networks—with an Erratum Note, German National Research Center for Information Technology GMD* Technical Report No. 148, 2001, p. 13.

[4] W. Maass, T. Natschläger, and H. Markram, *Real-Time Computing without Stable States: A New Framework for Neural Computation Based on Perturbations*, Neural Comput. **14**, 2531 (2002).

[5] G. Tanaka, T. Yamane, J. B. Héroux, R. Nakane, N. Kanazawa, S. Takeda, H. Numata, D. Nakano, and A. Hirose, *Recent Advances in Physical Reservoir Computing: A Review*, Neural Netw. **115**, 100 (2019).

[6] P. Antonik, A. Smerieri, F. Duport, M. Haelterman, and S. Massar, *FPGA Implementation of Reservoir Computing with Online Learning*, in *Proceedings of the 24th Belgian-Dutch Conference on Machine Learning* (2015), https://pdfs.semanticscholar.org/1fdf/965c77a08802276706c872f19073b9d3a0b1.pdf.

[7] C. Donahue, C. Merkel, Q. Saleh, L. Dolgovs, Y. K. Ooi, D. Kudithipudi, and B. Wysocki, *Design and Analysis of Neuromemristive Echo State Networks with Limited-Precision Synapses*, in *Proceedings of the 2015 IEEE Symposium on Computational Intelligence for Security*

*and Defense Applications (CISDA)* (IEEE, New York, 2015), pp. 1–6.

[8] M. Dale, J. F. Miller, S. Stepney, and M. A. Trefzer, *Evolving Carbon Nanotube Reservoir Computers*, in *Proceedings of the International Conference on Unconventional Computation and Natural Computation* (Springer, New York, 2016), pp. 49–61.

[9] C. Fernando and S. Sojakka, *Pattern Recognition in a Bucket*, in *Proceedings of the European Conference on Artificial Life* (Springer, New York, 2003), pp. 588–597.

[10] S. Ghosh, A. Opala, M. Matuszewski, T. Paterek, and T. C. Liew, *Quantum Reservoir Processing*, npj Quantum Inf. **5**, 35 (2019).

[11] J. Moon, W. Ma, J. H. Shin, F. Cai, C. Du, S. H. Lee, and W. D. Lu, *Temporal Data Classification and Forecasting Using a Memristor-Based Reservoir Computing System*, National electronics review **2**, 480 (2019).

[12] G. Van der Sande, D. Brunner, and M. C. Soriano, *Advances in Photonic Reservoir Computing*, Nanophotonics **6**, 561 (2017).

[13] L. Appeltant, M. C. Soriano, G. Van der Sande, J. Danckaert, S. Massar, J. Dambre, B. Schrauwen, C. R. Mirasso, and I. Fischer, *Information Processing Using a Single Dynamical Node as Complex System*, Nat. Commun. **2**, 468 (2011).

[14] Y. Paquot, J. Dambre, B. Schrauwen, M. Haelterman, and S. Massar, *Reservoir Computing: A Photonic Neural Network for Information Processing*, in *Nonlinear Optics and Applications IV* (International Society for Optics and Photonics, Bellingham, WA, 2010), Vol. 7728, p. 77280B.

[15] Y. Paquot, F. Duport, A. Smerieri, J. Dambre, B. Schrauwen, M. Haelterman, and S. Massar, *Optoelectronic Reservoir Computing*, Sci. Rep. **2**, 287 (2012).

[16] L. Larger, M. C. Soriano, D. Brunner, L. Appeltant, J. M. Gutiérrez, L. Pesquera, C. R. Mirasso, and I. Fischer, *Photonic Information Processing beyond Turing: An Optoelectronic Implementation of Reservoir Computing*, Opt. Express **20**, 3241 (2012).

[17] D. Brunner, B. Penkovsky, B. A. Marquez, M. Jacquot, I. Fischer, and L. Larger, *Tutorial: Photonic Neural Networks in Delay Systems*, J. Appl. Phys. **124**, 152004 (2018).

[18] J. D. Hart, L. Larger, T. E. Murphy, and R. Roy, *Delayed Dynamical Systems: Networks, Chimeras and Reservoir Computing*, Phil. Trans. R. Soc. A **377**, 20180123 (2019).

[19] F. Duport, A. Smerieri, A. Akrout, M. Haelterman, and S. Massar, *Fully Analogue Photonic Reservoir Computer*, Sci. Rep. **6**, 22381 (2016).

[20] F. Duport, A. Smerieri, A. Akrout, M. Haelterman, and S. Massar, *Virtualization of a Photonic Reservoir Computer*, J. Lightwave Technol. **34**, 2085 (2016).

[21] R. Martinenghi, S. Rybalko, M. Jacquot, Y. K. Chembo, and L. Larger, *Photonic Nonlinear Transient Computing with Multiple-Delay Wavelength Dynamics*, Phys. Rev. Lett. **108**, 244101 (2012).

[22] S. Ortín, M. C. Soriano, L. Pesquera, D. Brunner, D. San-Martín, I. Fischer, C. Mirasso, and J. Gutiérrez, *A Unified Framework for Reservoir Computing and Extreme Learning Machines Based on a Single Time-Delayed Neuron*, Sci. Rep. **5**, 14945 (2015).

[23] Q. Vinckier, F. Duport, A. Smerieri, K. Vandoorne, P. Bienstman, M. Haelterman, and S. Massar, *High-Performance Photonic Reservoir Computer Based on a Coherently Driven Passive Cavity*, Optica **2**, 438 (2015).

[24] B. Schneider, J. Dambre, and P. Bienstman, *Using Digital Masks to Enhance the Bandwidth Tolerance and Improve the Performance of On-Chip Reservoir Computing Systems*, IEEE Trans. Neural Netw. Learn. Systems **27**, 2748 (2015).

[25] D. Brunner, M. C. Soriano, C. R. Mirasso, and I. Fischer, *Parallel Photonic Information Processing at Gigabyte per Second Data Rates Using Transient States*, Nat. Commun. **4**, 1364 (2013).

[26] L. Larger, A. Baylón-Fuentes, R. Martinenghi, V. S. Udaltsov, Y. K. Chembo, and M. Jacquot, *High-Speed Photonic Reservoir Computing Using a Time-Delay-Based Architecture: Million Words per Second Classification*, Phys. Rev. X **7**, 011015 (2017).

[27] X. X. Guo, S. Y. Xiang, Y. H. Zhang, L. Lin, A. J. Wen, and Y. Hao, *Four-Channels Reservoir Computing Based on Polarization Dynamics in Mutually Coupled VCSELs System*, Opt. Express **27**, 23293 (2019).

[28] Y.-S. Hou, G.-Q. Xia, E. Jayaprasath, D.-Z. Yue, W.-Y. Yang, and Z.-M. Wu, *Prediction and Classification Performance of Reservoir Computing System Using Mutually Delay-Coupled Semiconductor Lasers*, Opt. Commun. **433**, 215 (2019).

[29] S. Ortín and L. Pesquera, *Reservoir Computing with an Ensemble of Time-Delay Reservoirs*, Cognit. Comput. **9**, 327 (2017).

[30] L. Keuninckx, J. Danckaert, and G. Van der Sande, *Real-Time Audio Processing with a Cascade of Discrete-Time Delay Line-Based Reservoir Computers*, Cognit. Comput. **9**, 315 (2017).

[31] B. Penkovsky, X. Porte, M. Jacquot, L. Larger, and D. Brunner, *Coupled Nonlinear Delay Systems as Deep Convolutional Neural Networks*, Phys. Rev. Lett. **123**, 054101 (2019).

[32] M. R. Salehi and L. Dehyadegari, *Optical Signal Processing Using Photonic Reservoir Computing*, J. Mod. Opt. **61**, 1442 (2014).

[33] K. Vandoorne, J. Dambre, D. Verstraeten, B. Schrauwen, and P. Bienstman, *Parallel Reservoir Computing Using Optical Amplifiers*, IEEE Trans. Neural Networks **22**, 1469 (2011).

[34] K. Vandoorne, W. Dierckx, B. Schrauwen, D. Verstraeten, R. Baets, P. Bienstman, and J. Van Campenhout, *Toward Optical Signal Processing Using Photonic Reservoir Computing*, Opt. Express **16**, 11182 (2008).

[35] K. Vandoorne, P. Mechet, T. Van Vaerenbergh, M. Fiers, G. Morthier, D. Verstraeten, B. Schrauwen, J. Dambre, and P. Bienstman, *Experimental Demonstration of Reservoir Computing on a Silicon Photonics Chip*, Nat. Commun. **5**, 3541 (2014).

[36] D. Brunner and I. Fischer, *Reconfigurable Semiconductor Laser Networks Based on Diffractive Coupling*, Opt. Lett. **40**, 3854 (2015).

[37] J. Bueno, S. Maktoobi, L. Froehly, I. Fischer, M. Jacquot, L. Larger, and D. Brunner, *Reinforcement Learning in a Large-Scale Photonic Recurrent Neural Network*, Optica **5**, 756 (2018).

[38] J. Dong, S. Gigan, F. Krzakala, and G. Wainrib, *Scaling up Echo-State Networks with Multiple Light Scattering*, in *Proceedings of the 2018 IEEE Statistical Signal Processing Workshop (SSP)* (IEEE, New York, 2018), pp. 448–452.

[39] J. Dong, M. Rafayelyan, F. Krzakala, and S. Gigan, *Optical Reservoir Computing Using Multiple Light Scattering for Chaotic Systems Prediction*, IEEE J. Sel. Top. Quantum Electron. **26**, 1 (2019).

[40] P. Antonik, N. Marsal, and D. Rontani, *Large-Scale Spatiotemporal Photonic Reservoir Computer for Image Classification*, IEEE J. Sel. Top. Quantum Electron. **26**, 1 (2019).

[41] U. Paudel, M. Luengo-Kovac, J. Pilawa, T. J. Shaw, and G. C. Valley, *Classification of Time-Domain Waveforms Using a Speckle-Based Optical Reservoir Computer*, Opt. Express **28**, 1225 (2020).

[42] S. Rotter and S. Gigan, *Light Fields in Complex Media: Mesoscopic Scattering Meets Wave Control*, Rev. Mod. Phys. **89**, 015005 (2017).

[43] H. Yu, K. Lee, and Y. Park, *Ultrahigh Enhancement of Light Focusing through Disordered Media Controlled by Megapixel Modes*, Opt. Express **25**, 8036 (2017).

[44] J. Pathak, B. Hunt, M. Girvan, Z. Lu, and E. Ott, *Model-Free Prediction of Large Spatiotemporally Chaotic Systems from Data: A Reservoir Computing Approach*, Phys. Rev. Lett. **120**, 024102 (2018).

[45] A. Saade, F. Caltagirone, I. Carron, L. Daudet, A. Drémeau, S. Gigan, and F. Krzakala, *Random Projections through Multiple Optical Scattering: Approximating Kernels at the Speed of Light*, in *Proceedings of the 2016 IEEE International Conference on Acoustics, Speech and Signal Processing (ICASSP)* (IEEE, New York, 2016), pp. 6215–6219.

[46] S. M. Popoff, G. Lerosey, R. Carminati, M. Fink, A. C. Boccara, and S. Gigan, *Measuring the Transmission Matrix in Optics: An Approach to the Study and Control of Light Propagation in Disordered Media*, Phys. Rev. Lett. **104**, 100601 (2010).

[47] N. Bertschinger and T. Natschläger, *Real-Time Computation at the Edge of Chaos in Recurrent Neural Networks*, Neural Comput. **16**, 1413 (2004).

[48] J. Pathak, Z. Lu, B. R. Hunt, M. Girvan, and E. Ott, *Using Machine Learning to Replicate Chaotic Attractors and Calculate Lyapunov Exponents from Data*, Chaos **27**, 121102 (2017).

[49] F. Schürmann, K. Meier, and J. Schemmel, *Edge of Chaos Computation in Mixed-Mode VLSI—A Hard Liquid*, in *Advances in Neural Information Processing Systems* (2005), pp. 1201–1208, http://papers.nips.cc/paper/2562-edge-of-chaos-computation-in-mixed-mode-vlsi-a-hard-liquid.pdf.

[50] J. Isensee, G. Datseris, and U. Parlitz, *Predicting Spatio-Temporal Time Series Using Dimension Reduced Local States*, J. Nonlinear Sci. **30**, 713 (2020).

[51] Dell Precision 7920 Workstation Desktop Tower with 2× Intel Xeon Gold 5120 (14 cores, 2.2–3.7 GHz Turbo, 19 Mb cache), 64 Gb 2666 MHz DDR4, 2 SSD M.2 2 Tb Pcie NVMe.

[52] Nvidia Tesla V100 32 GB configuration from Google Cloud Platform.

[53] A. Goetschy and A. D. Stone, *Filtering Random Matrices: The Effect of Incomplete Channel Control in Multiple Scattering*, Phys. Rev. Lett. **111**, 063901 (2013).

[54] D. Akbulut, T. Strudley, J. Bertolotti, E. P. A. M. Bakkers, A. Lagendijk, O. L. Muskens, W. L. Vos, and A. P. Mosk, *Optical Transmission Matrix as a Probe of the Photonic Strength*, Phys. Rev. A **94**, 043817 (2016).

[55] N. Keriven, D. Garreau, and I. Poli, *Newma: A New Method for Scalable Model-Free Online Change-Point Detection*, IEEE Trans. Signal Process. **68**, 3515 (2020).

[56] S. Maktoobi, L. Froehly, L. Andreoli, X. Porte, M. Jacquot, L. Larger, and D. Brunner, *Diffractive Coupling for Photonic Networks: How Big Can We Go?*, IEEE J. Sel. Top. Quantum Electron. **26**, 1 (2019).

[57] R. Ohana, J. Wacker, J. Dong, S. Marmin, F. Krzakala, M. Filippone, and L. Daudet, *Kernel Computations from Large-Scale Random Features Obtained by Optical Processing Units*, in *Proceedings of the 2020 IEEE International Conference on Acoustics, Speech and Signal Processing (ICASSP)* (IEEE, New York, 2020), pp. 9294–9298.

[58] T. W. Hughes, I. A. Williamson, M. Minkov, and S. Fan, *Wave Physics as an Analog Recurrent Neural Network*, Sci. Adv. **5**, eaay6946 (2019).

[59] G. Marcucci, D. Pierangeli, and C. Conti, *Theory of Neuromorphic Computing by Waves: Machine Learning by Rogue Waves, Dispersive Shocks, and Solitons*, Phys. Rev. Lett. **125**, 093901 (2020).

[60] Y. Zuo, B. Li, Y. Zhao, Y. Jiang, Y.-C. Chen, P. Chen, G.-B. Jo, J. Liu, and S. Du, *All-Optical Neural Network with Nonlinear Activation Functions*, Optica **6**, 1132 (2019).

[61] T. Yan, J. Wu, T. Zhou, H. Xie, F. Xu, J. Fan, L. Fang, X. Lin, and Q. Dai, *Fourier-Space Diffractive Deep Neural Network*, Phys. Rev. Lett. **123**, 023901 (2019).

[62] X. Guo, T. D. Barrett, Z. M. Wang, and A. Lvovsky, *End-to-End Optical Backpropagation for Training Neural Networks*, arXiv:1912.12256.

[63] https://github.com/jon-dong/reservoir-computing-python.

[64] J. W. Goodman, *Speckle Phenomena in Optics: Theory and Applications* (Roberts and Company, Greenwood Village, CO, 2007).

[65] P. Hohenberg and B. I. Shraiman, *Chaotic Behavior of an Extended System*, Physica (Amsterdam) **37D**, 109 (1989).

[66] A.-K. Kassam and L. N. Trefethen, *Fourth-Order Time-Stepping for Stiff PDEs*, SIAM J. Sci. Comput. **26**, 1214 (2005).

[67] R. A. Edson, J. E. Bunder, T. W. Mattner, and A. J. Roberts, *Lyapunov Exponents of the Kuramoto-Sivashinsky PDE*, ANZIAM J. **61**, 270 (2019).